\documentstyle[12pt,epsfig]{article}
\newcommand{\be}{\begin{eqnarray}}
\newcommand{\ee}{\end{eqnarray}}
\newcommand{\BE}{\begin{equation}}
\newcommand{\EE}{\end{equation}}
\begin{document}
\mbox{}
\rightline{December 1998}\\
\vspace{3.5cm}
\begin{center} 
%%%%%%%%%%%%%%%%%%%%%%%%%%%%%%%%%%%%%%%%%%%%%%%%%%%%%%%%%%%%%%%%%%%%%%%%%%
{\Large 
\bf A tracker solution to the cold dark matter cosmic coincidence problem}\\
\vspace{.5cm}
{\bf Ivaylo Zlatev$^{a,b}$ and Paul J. Steinhardt$^{a}$ 
}\\[.5cm]
{
$^a$Department of Physics, Princeton University, Princeton, NJ 08540
 \vspace{.1in} \\
$^b$Department of Physics and Astronomy,
University of Pennsylvania,
Philadelphia, PA 19104
}
\end{center}
\vspace{.5cm}

%%%%%%%%%%%%%%%%%%%%%%%%%%%%%%%%%%%%%%%%%%%%%%%%%%%%%%%%%%%%%%%%%%%%%%%%%%
\begin{abstract}
\noindent
Recently, we introduced the notion of ``tracker fields," a form of quintessence 
which has an attractor-like solution. Using this concept, we showed how to 
construct models in which the ratio of quintessence to matter densities today is 
independent of initial conditions. Here we apply the same idea to the 
standard cold dark matter component 
in cases where it is composed of oscillating fields. Combining these 
ideas, we can construct a model in which quintessence, cold dark matter,
and ordinary matter all contribute comparable amounts to the total energy 
density today irrespective of initial conditions. 
\end{abstract}
%\pacs{PACS number(s): 98.80.-k,98.70.Vc,98.65.Dx,98.80.Cq}
%]
\newpage
\noindent

One of the leading cosmological models nowadays is $\Lambda$CDM
which consists of a mixture of vacuum energy or cosmological constant
($\Lambda$) and cold dark matter (CDM).  
A number of recent observations suggest that $\Omega_m$,
the ratio of the (baryonic plus dark) matter density to the critical density, 
is significantly less than unity, and at the same time  recent 
supernova results suggest that the expansion of the Universe is 
accelerating \cite{supernova}. 
A serious problem with the $\Lambda$CDM scenario is the  ``cosmic coincidence 
problem'' \cite{zlatev1}.  Since the vacuum density is constant
and the matter density decreases as  the universe expands,
it appears that their ratio must be set to a specific, infinitesimal
value ($10^{-120}$) in the very early universe in order for the two densities to nearly
coincide today, some 15 billion years later. We will refer to the coincidence 
problem also as the ``initial conditions problem'' from now on.
Recently,  we proposed the notion of ``tracker quintessence fields" to resolve the 
$\Lambda$ coincidence problem \cite{zlatev1}.
Quintessence, as a substitute for the cosmological constant, is a slowly-varying,
spatially inhomogeneous component with a negative equation of state.\cite{Cald98}
An example of quintessence is the energy associated with a scalar field ($Q$)
slowly evolving down its potential $V(Q)$.\cite{Ratra,Friem,Cald98,Ferreira}
``Tracker fields" are a form of quintessence in which the tracker field
$Q$ rolls down a potential $V(Q)$ according to an attractor-like
solution to the equations-of-motion.  The tracker solution is an attractor in
the sense that a very wide range of initial conditions for $Q$ and $\dot{Q}$
rapidly approach a common evolutionary track, so that the cosmology is
insensitive to the initial conditions.  Tracking has an advantage similar to
inflation in that a wide range of initial conditions is funneled into the same
final condition.

In the present Letter we examine the initial conditions problem for a class
of theories that treat CDM as composed of oscillating fields. There are 
two components that determine $\Omega_{CDM}$: the value of the field 
$\psi$ and its effective mass $m_{eff} \sim V''(\psi)$ ($'$ is a derivative with respect to
the field $\psi$). The 
density of this CDM candidate then is described by 
$\rho_{CDM} \sim m_{eff}^2 \psi^2$. 
The axion field is an example of an oscillating field CDM candidate.
Once the oscillations begin, the density redshifts, just like $CDM$, 
as one over the scale factor cubed. 
The initial conditions problem for the oscillating field model of $CDM$
is that there is one {\em unique} initial density or,
equivalently a unique value of $\psi$, that leads to the present day
observed CDM density. If one imagines that $\psi$ is set randomly, or by
equipartition after inflation, the probability of obtaining the right density
today is infinitesimal. 
 
In this Letter we propose a resolution of the initial conditions problem based 
on applying  the ``tracker fields" idea to models of CDM composed 
of oscillating fields. 
To this end we will construct potentials that have attractor-like solutions 
at early times, but at later times the solutions are oscillatory and CDM-like. 
Thus, at the beginning, the potential funnels a 
large number of different initial conditions into one state. At late times
this state enters an oscillatory phase and the field behaves as the dark matter 
oscillatory candidates discussed above. The removal of the initial conditions 
dependence is achieved at the expense of introducing an additional tuned parameter 
in the CDM field potential. 
We also  point out how, by combining 
Quintessence with CDM tracker solutions, one can construct a toy model in which
the density ratios today are all determined by parameters in the potential and
are insensitive to initial conditions. The possibility that the initial 
energy densities of all cosmic density constituents were in equilibrium initially is 
allowed for. 

Before we embark on a discussion of oscillatory fields as CDM candidates, 
we should first answer the question: what observations characterize CDM and 
what field potentials satisfy these observations. The two observations that
characterize CDM are (1) equation of state of zero since CDM is non-relativistic
and (2) CDM should cluster on scales larger than a $Mpc$ in order for galaxies
and quasars to form at a moderate redshift. There are a number of potentials 
that yield equations-of-state equal to zero. We will separate them in two categories.
The first one is the category of potentials with either constant mass or a 
variable mass
that became  bigger than the Hubble parameter at some time in the past. As soon as 
the Hubble parameter redshifts to a value several times smaller than the field mass,
the field begins to oscillate. The quadratic potential $m^2\psi^2$ is the most widely 
considered example of this category. The second category is comprised of 
potentials that have effective masses that decrease at the same rate as 
the Hubble parameter and never become bigger than it. A widely studied example
of the second category is the exponential potential $Ae^{\alpha \psi}$ 
\cite{Ferreira}. 
It is the second observation, the necessity for CDM to cluster above a $Mpc$
scale that renders only the first category as valid CDM candidates. 
This is so because the field clusters like CDM only on scales larger than 
the mass of the field \cite{Friem,quadr_nrpert}.
On smaller scales cluster formation is suppressed. 
By expanding the analysis initiated in \cite{Friem} one can show that 
for galaxies and quasars to
form at a moderate redshift, the mass of the field has to be larger than 
$4-20\times 10^{-36} GeV$ which in turn translates in $z > (1-3.5)\times 10^4$
as the redshift at which the oscillations can start at the latest.

The value of $\Omega_{cdm}$ obtained today is highly sensitive to the
initial condition for the oscillating field, $\psi_i$.  Let us consider
the quadratic potential, $m^2 \psi^2$,  for example.
The values of $m$ and $\psi_i$ are fixed by the condition that 
$\Omega_{\psi}$ is equal to the observed value of $\Omega_{cdm}$ today.
Although the field is presumed to be oscillating today (which requires
$m> 3H$), it was frozen at some $\psi_i$ for most of its history
when $m \ll H$. Hence, we only need consider the initial expectation 
value and not the initial kinetic energy of $\psi_i$.  
If  $\psi$ where set to $\psi_i$ at the end
of inflation, say, then it would maintain that same value until 
very late in the history of the universe when $H$ decreases below $m$.
However, this requires that 
$\rho_\psi$  be set many orders of magnitude less than the matter-radiation 
energy at the end of inflation. 
A different initial value of $\rho_{\psi}$ leads to a different $CDM$
density today and only a limited range is compatible with large-scale
structure formaetion.

One of the most widely discussed oscillating field CDM candidates is the axion. The axion field
is a quantum field $\psi$ added to particle-physics models in order to solve the strong 
CP problem \cite{quinn}.  
It has been shown that 
$\Omega^0_a h^2 \simeq 10^7 (f_a/M_p)(\psi_i/f_a)$ \cite{wilcheck}  where by
subscript $i$ we mean the value of the quantity when the harmonic 
oscillations of the field begin, $f_a$ is the Pecci-Quinn symmetry
breaking scale and $\Omega^0_a$ is the fraction of the present 
cosmic energy density that is attributed to the axions. 
The axion field acquires a high enough mass to commence oscillations
at around the $QCD$ scale of $200 MeV$ (before that it is effectively massless). 
Although it is commonly assumed that the initial value of $\psi$ is $\psi_i \sim f_a$ 
and this is used to estimate $\Omega_a$, in fact, $\Omega_a$ is very sensitive to the precise
value of $\psi_i$ and can take on any value between $0$ and $1$ (assuming a flat universe say).
The initial value of $\psi$ has to be set very precisely (and spatially uniformly) to obtain
the correct $\Omega_a$ today. The initial condition corresponds to setting the initial 
value of $\rho_{axion}$ over $70$ orders of magnitude less than the energy density at the end of
inflation (when $\psi_i$ was presumably set).   
Some discount this tuning as being no different from fixing any other
parameter in a theory, such as Newton's constant. If this be the case
for  you, dear reader, there is no point to continuing this paper 
since you do not acknowledge the problem we attempt to address.
Others (including us)
would object that the situation is not analogous to the case of 
measuring Newton's constant:
By the structure of the theory, 
$G$ must take some value and measuring it through 
gravitational effects is reasonable;
it seems remarkable that the 
axion, invented to solve the $U(1)$ problem, should happen to have the
right coupling and initial conditions to comprise the dark matter today.
Bear in mind that the existence of stars, galaxies and large-scale structure
depend on the $CDM$ density being neither too large nor too small,
within a fairly narrow range.  The high sensitivity to initial conditions
seems particularly disturbing since they are determined by some 
random process, most likely.

While we proposed the tracking mechanism in order to resolve the $\Lambda$ initial
conditions problem, we will now apply it to resolve the CDM initial conditions problem as well.
Let us first review the essence of tracking. Tracker fields have an equation of 
motion with attractor-like solutions in which a very wide range of initial conditions 
rapidly converge to a common, cosmic evolutionary track. The central theorem that
we proved was that tracking behavior with $w_Q < w_B$ occurs for any potential in which 
$\Gamma \equiv V''V/(V')^2 > 1$ and is nearly constant ($|d(\Gamma-1)/Hdt| \ll |\Gamma-1|$).
We needed $w_Q < w_B$ in order to explain the present day acceleration of the 
Universe suggested by the recent supernova results. While looking for trackers with 
quintessential behavior ($w_Q < w_B$), we discovered another category of trackers
with  $w_B \leq w_Q < (1/2)(1+w_B)$  provided $\Gamma \leq 1$ and nearly constant.
While these trackers do not have the accelerating properties of Quintessence,
they are useful for resolving the CDM initial conditions problem.
Potentials that yield such behavior are for example $V(\psi) = Ae^{\alpha \psi},~
V(\psi) = B \psi^\beta;~\beta > 10$. 
Since these potentials exhibit tracking behavior,
they funnel a large number of initial conditions in one final state. If these 
potentials also had the CDM clustering properties, then they would have been the 
resolutions of the CDM initial conditions problem. Alas, 
these potentials by themselves cannot play the
role of oscillating field CDM since their effective masses ($V''$) decrease at the 
same rate as the 
Hubble parameter does and thus fail to reproduce the observed clustering at scales
larger than a $Mpc$. Thus, we do not want the tracker condition satisfied at all times.
We want it satisfied at early times so that a large number of initial conditions are 
funneled in one final state. At late times we want the field to exhibit 
non-tracking, oscillatory clustering behavior. 
One way to achieve this goal would be to add a tracker potential to
the quadratic one: $V=m^2 \psi^2 + V_{tracking}$ and to adjust the coefficient in 
$V_{tracking}$ in such a way that $V_{tracking}$ dominates at early times but is 
sub-dominant at late times. This way we gain independence of initial conditions 
at the price of tuning a parameter in the potential. 

As a pedagogical illustration, we will first consider the toy model 
\be
         V= m^2 \psi^2 + \beta \psi^\alpha ;~~~\alpha > 10
\ee
where $\beta$ has the dimensions of $M_p^{4-\alpha}$ 
and for calculational purposes we have taken $\alpha=12$. 
The mass of the field $m$ is determined by the minimum scale at which we want our 
CDM to exhibit clustering properties.  If only the quadratic term were present,
then there would have been only one unique initial value of the field ($\psi_i$) 
that that would have led to present day CDM energy density. This unique initial
value is denoted by a solid circle in Fig. 1. In order to resolve this 
initial conditions problem, we add an additional term to the potential that  
is equal to the quadratic term at the time the oscillations begin; this term 
dominates at earlier
times and is sub-dominant at late times. This is achieved by tuning the parameter
$\beta$. This additional term exhibits tracking behavior as already discussed;
but it exhibits it only at early times when it is dominant. At late times 
it is sub-dominant and the field can cluster. Namely, we have taken a potential 
where before we had to set $\psi$ by hand to get the right value and now
we add a term that dominates at early times and automatically drives $\psi$ to the 
right value.  As shown in Fig. 1 while tracking at early times, the potential 
funnels a large range (spanning over seventy orders of magnitude) 
of initial conditions in one final state. 

\begin{figure}
 \epsfxsize=3.3 in \epsfbox{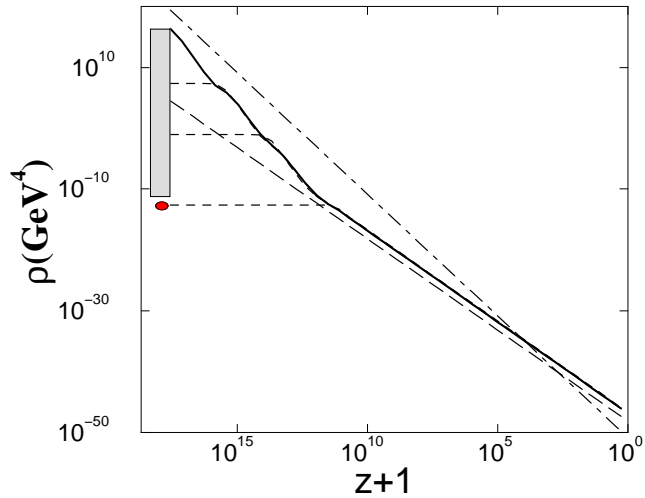}
\caption{ The potential is $V= m^2 \psi^2 + \beta^2\psi^{12}$. The choice of 
$z=10^{18}$(rather than inflation) has been arbitrarily chosen as the initial time for 
convenience of computation and illustration and we have used $\Omega_{CDM}=0.95$ today.
The bar on the left hand side represents the range of initial values (spanning over
70 orders of magnitude if we extrapolate back to inflation) that converge to a 
$\Omega_{CDM}=0.95$ today. The solid circle represents the unique choice of initial 
conditions that gives $\Omega_{CDM}=0.95$ today if the CDM potential is 
pure quadratic $V= m^2 \psi^2 $.  The
dot-dashed line is the radiation density, the long dashed is the baryon density,
the solid thick line is the tracker solution and the short dashed lines are
examples of different initial conditions that undershoot the tracker
solution but converge on it.
}
\end{figure}

We would like to emphasize that our general approach does not depend on the 
specific choice of potential. We could have achieved the same desirable effect
of funneling a large number of initial conditions in one final state via 
any combination of quadratic with any number of higher power polynomial terms.
While in an arbitrary combination of numerous terms one would have to adjust 
a number of parameters, a gaussian-like potential is a particular combination
of infinitely many high power polynomial terms with only two parameters.
As shown in Fig. 2, we can resolved the CDM 
initial conditions problem using a potential of the type
\be
        V = A [e^{\lambda \psi^2 }-1].
\ee
Expanding this potential in a Taylor series, we see that at early times 
the series is dominated by the high power polynomial terms which exhibit 
tracking behavior as argued above, but at late times the quadratic term
dominates and the field starts to oscillate and cluster like CDM. Using the 
tracker equations of motion \cite{zlatev_2} one can show rigorously
that $\lambda$ is set by $\Omega_\psi^0$ and $\lambda A$, being the effective mass
of the oscillating field, determines the redshift at which the oscillations 
commence. 
Hence, two observable quantities determine the two potential parameters.
This example is only a toy model to illustrate that the 
scenario is dependent on only two parameters.  In general,
polynomial and exponential 
 (non-perturbative) potentials with high-order powers of the field
are considered in supersymmetric particle physics theories \cite{binetr}.

\begin{figure}[t]
\centering
\epsfig{file=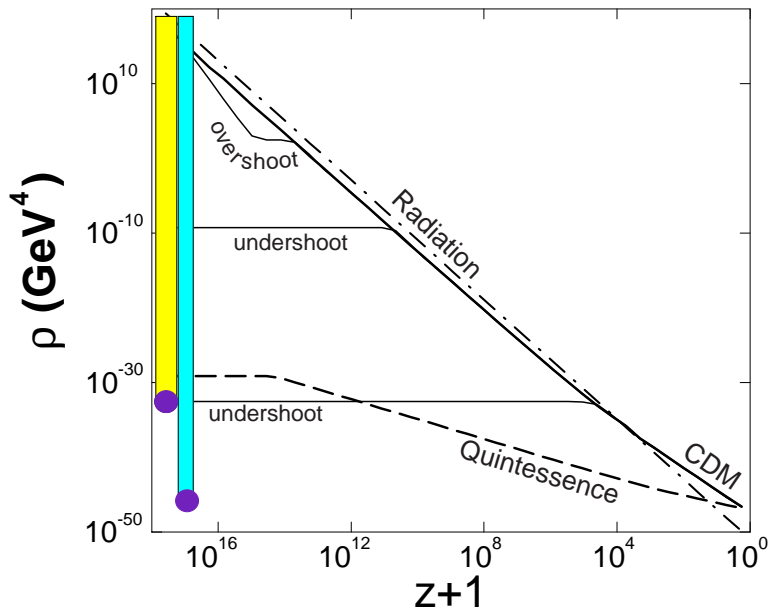,height=10cm}
\caption{ The potential is $V= Ae^{\lambda \psi^2} + B/Q$ where
dot-dashed line is the radiation density, the long dashed is the tracker $Q$ density
$\Omega^0_Q \sim 0.68$,
the solid thick lines are the tracker cdm density $\Omega^0_\psi \sim 0.27$; thick - tracker,
thin over and under-shoots that converge on the cdm tracker. 
The solid circles represents the unique choices of initial
conditions that gives $\Omega_{CDM}=0.27$ and $\Omega^0_Q \sim 0.68$ today if 
the CDM potential is pure quadratic $V= m^2 \psi^2 $ and we have cosmological constant
instead of tracking quintessence. The shaded bars on the left show how the tracker 
idea dramatically increases the range of viable initial conditions.
}
\label{fig2}
\end{figure}

Current data suggests that the universe contains both a CDM component and a 
$\Lambda$ or Quintessence component. Arranging either component to be 
comparable to the baryon density today appears to require a fine tuning
of initial conditions (if the CDM component never reached thermal equilibrium).
It is interesting that both fine tunings can be resolved by a common 
mechanism, tracking. A trivial example is the combined two-field
potential
\be
	V = Ae^{\lambda \psi^2} + B/Q^\alpha
\ee
where $Q$ stands for Quintessence (see Fig. 2). The reaction of 
$\Omega_Q$ and $\Omega_{CDM}$ to $\Omega_{baryon}$ does not depend on the 
initial conditions of the components or the initial values of $Q$, $\dot Q$,
$\psi$ and $\dot\psi$  but only on the parameters of the potential. 
Although this example is artificial, it points out the power of tracking 
to resolve problems of initial conditions in cosmology and offers hope of an
even more economical model.

%%%%%%%%%%%%%%%%%%%%%%%%%%%%%%%%%%%%%%%%%%%%%%%%%%%%%%%%%%%%%%%%%%%%%%%%%%

\begin{thebibliography}{99}

\bibitem{supernova} S. Perlmutter, {\it et al}, astro-ph/9712212;
A. G. Riess, {\it et al}, astro-ph/9805201.

\bibitem{zlatev1} I. Zlatev, L. Wang and P. J. Steinhardt,
Phys. Rev. Lett. {\bf 82} (1999) 896.

\bibitem{Cald98} R.R. Caldwell, R. Dave and P.J. Steinhardt,
Phys. Rev. Lett.  {\bf 80}  (1998) 1582.

\bibitem{Ratra}
B. Ratra, and P.J.E. Peebles, Phys. Rev. D {\bf 37} (1988) 3406;
P.J.E. Peebles and B. Ratra,  ApJ {\bf 325} (1988) L17.

\bibitem{Friem}
J.A. Frieman, {\it et al.}  Phys. Rev. Lett. {\bf 75} (1995) 2077.

\bibitem{Ferreira}
P.G. Ferreira and M. Joyce,  Phys. Rev. Lett. {\bf 79} (1997) 4740;
Phys. Rev. D {\bf 58}(1998) 023503.

\bibitem{quadr_nrpert}R. Brandenberger, Phys. Rev. D {\bf 32} (1985) 501;
M. Klopov, B. Malomed and Ya. Zeldovich, Mon. Not. R. asrt. Soc. {\bf 215}
(1985) 575.

\bibitem{quadr} M. Turner, Phys. Rev. D {\bf 28} (1983) 1243.
J. McDonald, ibid {\bf 48} (1993) 2573.

\bibitem{zlatev_unp}I. Zlatev, R. Caldwell, L. Wang, P. J. Steinhardt (in preparation).

\bibitem{mt_axionsearch}M. Turner,  Phys. Rep. {\bf 197} (1990) 67.



\bibitem{quinn} R. Peccei and H. Quinn, Phys. Rev. Lett. {\bf38} (1977) 1440.
 
\bibitem{wilcheck} J. Preskill, M. Wise and F. Wilczeck, Phys Letters
{\bf 120 B} (1983) 127.

\bibitem{zlatev_2} P. J. Steinhardt, L. Wang and I. Zlatev (astro-ph/9812313).

\bibitem{binetr}P. Binetruy, M. Gaillard, Y. Wu, Phys. Lett. {\bf B412}
(1997) 288.




%\bibitem{turner93}M. Turner, astro-ph/9302003 and references therein. 

%\bibitem{quadr_dicke}P. J. Steinhardt and C. Will, Phys. Rev. D {\bf 52}
%(1995) 628; J. McDonald, Phys. Rev. D {\bf 44} (1991) 2325;
%ibid {\bf 48} (1993) 2462.

%\bibitem{quadr_rpert}B. Ratra,  Phys. Rev. D {\bf 44} (1991) 352;
%J. Hwang (astro-ph/9610042).
 
%\bibitem{wang}L. Wang, R. Caldwell, J. Ostriker and  P. J. Steinhardt (astro-ph/9901388).

%\bibitem{campbell} B. A. Campbell, A. Linde and K. A. Olive, Nucl. Phys. 
%{\bf B366} (1991) 146.

%\bibitem{brus} R. Brustein and P. J. Steinhardt, Phys. Lett. {\bf B302} 
%(1993) 196. 

\end{thebibliography}
\end{document}